\documentclass[preprint]{JHEP} 
\usepackage{epsfig}
\usepackage{amsmath}
\usepackage{amssymb,amsfonts}

\def\bs{\begin{subequations}}
\def\es{\end{subequations}}

\def\sp{\;\;\;,\;\;\;}

\newcommand{\irrep}[1]{\ensuremath{\boldsymbol{#1}}}

\newcommand{\D}{{\cal D}}

\newcommand{\M}{{\cal M}}

\newcommand{\Zint}{\mathbb{Z}}
\newcommand{\Real}{\mathbb{R}}

\newcommand{\lh}{\ensuremath{l_{\rm H}}}

\newcommand{\lii}{\ensuremath{l_{\rm II}}}

\newcommand{\lp}{\ensuremath{l_{\rm P}}}
\newcommand{\gh}{\ensuremath{g_{\rm H}}}
\newcommand{\gii}{\ensuremath{g_{\rm II}}}

\newcommand{\giis}{\ensuremath{g_{\rm 6IIA}}}
\newcommand{\giit}{\ensuremath{g_{\rm 3IIA}}}

\newcommand{\ghs}{\ensuremath{g_{\rm 6H}}}
\newcommand{\ght}{\ensuremath{g_{\rm 3H}}}
\newcommand{\ghf}{\ensuremath{g_{\rm 4H}}}

\newcommand{\ben}{\begin{enumerate}}
\newcommand{\een}{\end{enumerate}}
\newcommand{\iia}{{\rm IIA}}

\renewcommand{\sp}{\ ,\qquad}


\numberwithin{table}{section}

\def\e{\epsilon}

\def\e{\epsilon}
\def\be{\begin{equation}}
\def\ee{\end{equation}}
\def\p{\partial}

\title{Exact thresholds and instanton effects\\
in $D=3$ string theories}

\author{Niels A. Obers\thanks{
Work supported in part by TMR network ERBFMRX-CT96-0045.}\\ Nordita
and Niels Bohr Institute\\ Blegdamsvej 17, DK-2100 Copenhagen,
DENMARK\\ E-mail: \email{obers@nordita.dk}}
\author{Boris Pioline\footnote{On
leave of absence from LPTHE, Universit{\'e} Pierre et Marie Curie,
PARIS VI and Universit{\'e} Denis Diderot, PARIS VII, Bo\^{\i}te
126, Tour 16, 1$^{\it er}$ {\'e}tage, 4 place Jussieu, F-75252
Paris CEDEX 05, FRANCE}\\
Jefferson Physical Laboratory, Harvard University\\
Cambridge, MA 02138, USA\\
E-mail: \email{pioline@physics.harvard.edu}}

\preprint{\hepth{0006088}\\
NBI-HE-00-29 \\NORDITA-2000/54 HE \\
HUTP-00/A022\\LPTHE-00-22}      

\abstract{Three-dimensional string theories with 16 supersymmetries
are believed to possess a non-perturbative U-duality symmetry
$SO(8,24,\Zint)$. By covariantizing the heterotic one-loop amplitude
under U-duality, we propose an exact expression for the
$(\partial \phi)^4$ amplitude, that reproduces known perturbative limits.
The weak-coupling expansion in either of the heterotic, type II or
type I descriptions exhibits the well-known instanton effects, plus
new contributions peculiar to three-dimensional theories,
including the Kaluza-Klein 5-brane, for which we extract the summation measure.
This letter is a post-scriptum to \hepth{0001083}.}

\keywords{Nonperturbative Effects, String Duality, p-branes} 

\begin{document}

\section{Introduction}
In a recent article \cite{Kiritsis:2000zi},
we have submitted the conjectured heterotic-type II
duality to its most detailed test to date, by showing the identity of the
heterotic one-loop--exact four--gauge-boson $F^4$ threshold in six dimensions
with the tree-level amplitude in type IIA compactified on $K_3$. This
test was carried out at the $\Zint_2$ orbifold point of $K_3$, and took
several non-trivial facts to succeed, including the invariance under
triality of automorphic forms of the T-duality group $SO(4,4,\Zint)$
\cite{Obers:1999um}.
As a spin-off of this analysis, we obtained exact amplitudes in type I',
F-theory and M-theory on $K_3$ at a singular point, as well as in type II
compactified on $K_3\times T^{d}$ at the orbifold point,
and extracted the contributions from D-brane instantons (for $d=1,2$)
and for the first time NS5-brane instantons ($d=2$).

In this letter, we extend our analysis to the case of the
heterotic string compactified on $T^7$ to $D=3$ dimensions, or
equivalently type II string compactified on $K_3\times T^3$. This
generalization is of interest for several reasons. Firstly,
three-dimensional gravity theories have an interesting infrared
behavior, due to the deficit angle created by massive particles,
which allows to have a supersymmetric vacuum without Bose-Fermi
degeneracy \cite{Witten:1995cg}. The possibility of dynamically
generating a fourth dimension may then offer a solution to the
cosmological constant problem \cite{Witten:1995rz}.
Three-dimensional gauge theories may also have non-trivial
infrared fixed points \cite{Banks:1997my} which are
far from being understood. Secondly, in contrast to $D>3$, the
perturbative moduli unify with the heterotic dilaton and the
dualized gauge fields into a simple
symmetric manifold $SO(8)\times SO(24)\backslash SO(8,24)$
\cite{Marcus:1983hb}, identified under the conjectured U-duality
symmetry $SO(8,24,\Zint)$ \cite{Sen:1995wr}. Such a unification
also happens in the case of M-theory compactified on a torus $T^d$
(see \cite{Obers:1998fb} for a review) where the use
of automorphic forms has proven efficient, and in type IIB on
$K_3$, where the U-duality $SO(5,21,\Zint)$ symmetry has been
put to little use (see however \cite{Kiritsis:2000zi} for
U-duality invariant $t_{12}H^4$ couplings, obtained by duality
from the heterotic string on $T^5$). Because of the unification
of the dilaton with other moduli, any U-duality invariant amplitude is thus
bound to contain instanton corrections along with the perturbative
contributions, and hence to give us information in particular
about the elusive heterotic instantons. Thirdly, string theories
in three dimensions possess new instanton configurations that were
not present in $D>3$, namely the Kaluza-Klein 5-brane 
instantons. These instantons are obtained by tensoring a Taub-NUT gravitational
instanton asymptotic to $\Real^3\times S^1$ with a flat $T^6$,
where $T^7=S^1\times T^6$. They are the ten-dimensional Euclidean version
\cite{Townsend:1995kk}
of the Kaluza-Klein monopoles introduced in
\cite{Sorkin:1983ns}. Their
worldvolume action has been constructed
in \cite{Bergshoeff:1997gy},
and their dynamics analyzed in \cite{Sen:1997zb}.
Finally, $D=3$ is just one step away from $D=2$, where
the U-duality group becomes infinite-dimensional, and predicts an
infinite set of particles with exotic dependence $1/g_s^{n+3}$ on
the coupling \cite{Elitzur:1997zn,Obers:1998fb}.
Similar states already appear as particles in $D=3$ with mass
$1/g_s^3$. Instantons in $D=3$ are however free of these infrared
problems, and the study of exact amplitudes in $D=3$ may shed
light on the non-perturbative spectrum.

The plan and main results of this paper are as follows. In Section
2 we recall the reader about the way $SO(8,24,\Zint)$ duality
arises in the heterotic string on $T^7$, and how it manifests
itself on the type IIA/$K_3\times T_3$ side. In particular, we
show how the type II $T^3$ moduli are related by triality to the
ones on the heterotic side, in much the same way as the type II
$K_3$ moduli are related by triality to the heterotic $T^4$ moduli
at the orbifold point \cite{Kiritsis:2000zi,Nahm:1999ps}. In
Section 3, we propose a U-duality invariant completion of the
heterotic one-loop $F^4$ amplitude, which also reproduces the type
II tree-level result at the orbifold point. From the
non-perturbative expansion on the heterotic side we are able to
identify new heterotic 5-brane and KK5-brane instanton effects.
The corresponding expansion on the type II side reveals, along
with the D$p$-brane and NS5-brane instanton effects found in
\cite{Kiritsis:2000zi}, D6-brane and new KK5-brane instanton
contributions as well. Finally, we comment on the type I dual theory
at the orbifold point, for which, besides the known perturbative
terms \cite{Tseytlin:1996fy} and
D1-instantons \cite{Bachas:1997mc,Kiritsis:1997hf,Foerger:1998kw}, we identify
D5-brane and KK5-brane instanton contributions.

\section{U-duality and heterotic-type II triality in $D=3$ \label{trial}}
In this section, we give a brief survey of the appearance of the
$SO(8,24,\Zint)$ U-duality symmetry in heterotic on $T^7$
or type II on $K_3\times T^3$. For simplicity, we consider mainly
compactifications on square tori without gauge background,
even though the full duality
becomes apparent only when including the Wilson line and vielbein moduli
on the heterotic side, and the Ramond backgrounds on the type II side.

\subsection{$SO(8,24)$ invariance in heterotic on $T^7$ \label{het7} }
The heterotic string compactified on $T^7$ possesses the well-known
T-duality invariance $SO(7,23,\Zint)$. The scalars fields coming from the
reduction of the ten-dimensional metric, B-field and $U(1)$ gauge fields
take value in a coset $[SO(7)\times SO(23)] \backslash SO(7,23)$,
parameterized as usual by a symmetric matrix $M_{7,23}$.
The three-dimensional coupling constant $1/\ght^2=V_7/(\gh^2 \lh^7)$
takes value in a separate $\Real^+$ factor, inert under T-duality.
In addition, there are $7+23$ $U(1)$ gauge fields, which in three
dimensions can be dualized into as many scalars. This is achieved by adding
a Lagrange multiplier term $\phi_a dF^a$ to the gauge kinetic term
$(\lh/\ght^2) F^a (M_{7,23})^{-1}_{ab} F^b$ and integrating out the field
strength $F^a$, which yields
\begin{equation}
\label{vsdua}
*F^a =  \frac{\ght^2}{\lh} M_{7,23}^{ab} d\phi_b \ . 
\end{equation}
Together with the dilaton and the $SO(7,23)$ moduli, these scalars
span an $[SO(8)\times SO(24)]\backslash SO(8,24)$ symmetric space,
which encompasses all bosonic fields except the graviton.

Another way of getting to this result is to first compactify the
heterotic string on $T^6$ to four dimensions, and then further
to $D=3$. In $D=4$, the moduli space has two factors,
one is the usual $[SO(6)\times SO(22)]\backslash SO(6,22)$
factor, acted upon by $SO(6,22,\Zint)$ T-duality, and the second
is $U(1)\backslash Sl(2)$, parameterized by the complex modulus
$S=B+i/\ghf^2$, where $1/\ghf^2=V_6/(\gh^2\lh^6)$ is the four-dimensional
string coupling constant and $B$ the scalar dual to the Neveu-Schwarz
two-form in four dimensions. The action of $Sl(2,\Zint)$ on $S$ is
conjectured to be a non-perturbative symmetry of the heterotic string
compactified on $T^6$.
Upon further compactification to $D=3$, the radius $R_7$, the
Wilson lines of the 6+22 gauge fields and their scalar dual
in $D=3$ make up again the $SO(8,24)$ coset space.

This last point of view allows to easily identify the subgroup
of $SO(8,24)$ which remains as a quantum symmetry. Indeed, the
action of the Weyl generator $\ghf\to 1/\ghf, \lh\to \ghf^2\lh$.
of the $Sl(2,\Zint)$ S-duality group translates in terms of
the three-dimensional variables into the exchange of $1/\ght^2$
and $R_7/\lh$. Using the Weyl group of $Sl(7)\subset SO(7,23)$,
we can transform $1/\ght^2$ into $R_i/\lh$ for any radius of $T^7$.
This implies that we can think of the $SO(8,24)$ scalars as
the moduli of an heterotic compactification on $T^7\times S^1$,
where the radius of $S_1$ is given by
\begin{equation}
\label{r8def}
R_8/\lh  = 1/\ght^2 \ . 
\end{equation}
This compact circle therefore appears as a dynamically generated dimension
decompactifying at weak coupling\footnote{It may seem more sensible to
define the T-dual radius $\hat R_8/\lh = \ght^2$ which decompactifies
at strong coupling, but the limits $R\to 0$ and $R\to \infty$ are
equivalent in the heterotic string.}. Note that this is {\it not} the usual
M-theory direction, whose radius is instead $R_{11}=\gh \lh$.
In particular, U-duality implies that there should be
enhanced gauge symmetry
at $R_8=\lh$, {\it i.e.} $\ght=1$.

We have therefore identified the dilatonic moduli parameterizing the
Abelian part of $SO(8,24)$,
\begin{equation}
\label{hetdil}
x_{i=1\dots 8}= ( R_1/\lh, \dots, R_7/\lh, 1/\ght^2) \ . 
\end{equation}
Including the 30 scalars $\phi_a$ of \eqref{vsdua}, the vielbein
parameterizing the $SO(8)\times SO(24)$ $\backslash SO(8,24)$
coset can be chosen as
\begin{equation}
\label{hetbein2} e_{8,24}=\begin{pmatrix} \ght^2 & & \\ & e_{7,23} & \\
& & \ght^{-2}   \end{pmatrix} \cdot
\begin{pmatrix} 1  & \phi  & -\phi ~\eta_{7,23}~ \phi^t/2   \\ & 1_{30} &
-\eta_{7,23} \phi^t  \\ & & 1
\end{pmatrix} \sp e_{8,24}^t~ \eta_{8,24}~ e_{8,24} = \eta_{8,24}
\end{equation}
where $e_{7,23}$ is the vielbein of the perturbative $SO(7,23)$ moduli
and we have defined
\begin{equation}
\label{eta}
\eta_{d,d+16}=\begin{pmatrix}
&&1_d \\&1_{16}&\\ 1_d &&\end{pmatrix} \ . 
\end{equation}

Having in mind the decompactification of our forthcoming three-dimensional
results to $D=4$, we now discuss in some more detail the embedding
of $SO(6,22)$ into $SO(8,24)$.
The 7th direction taking from to $D=4$ to $D=3$
and the non-perturbative 8th direction
form a dynamically generated torus $T^2$ with complex structure and
K\"ahler moduli
\begin{equation}
\label{t2mod}
U = B + i \frac{R_8}{R_7}=S \sp
T = A + i \frac{R_7 R_8}{\lh^2}=A+ i\frac{R_7^2 V_6 }{\gh^2 \lh^8} \ , 
\end{equation}
where we recognize the complex structure $U$ as the S modulus of the
four-dimensional theory, and the imaginary part of the K\"ahler modulus 
$T$ as the action of the Kaluza-Klein monopole on $T^7$.
The axionic part $A$ is the dual of the Kaluza-Klein gauge
field arising in the reduction of the metric from $D=4$ to $D=3$.
Altogether, the $SO(6,22)$ moduli of $T^6$ join the $SO(2,2)$ moduli
of the non-perturbative $T^2$ and the $2\times(6+22)$ Wilson lines
of the four-dimensional gauge fields and their duals, into the
$SO(8,24)$ coset.

To summarize, we have seen that the
$SO(8,24,\Zint)$ duality arises from a combination
of the $SO(7,23,\Zint)$ T-duality symmetry and the $Sl(2,\Zint)$
electric-magnetic S-duality in 4 dimensions: it should therefore be
an exact quantum symmetry of the three-dimensional theory.

\subsection{Heterotic-type II double triality}
The heterotic string compactified on $T^7$ and the type II string
compactified on $K_3\times T^3$ share the same supergravity action.
Instead of going through the same procedure as above, we shall
identify the $SO(8,24)$ symmetry on the type II side by using
heterotic-type II duality in six dimensions \cite{Sen:1995cj}.
The identification of the moduli in $D=6$ has been discussed in
great detail in \cite{Kiritsis:2000zi}, where it was shown that
the heterotic moduli are related to the type II ones by $SO(4,4)$
triality, mapping the vector representation to the conjugate
spinor.
At the $T^4/\Zint_2$ orbifold point with a square $T^4$,
this reduces to
\begin{equation}
R_1\vert_{\rm H}={\sqrt{R_1 R_2 R_3 R_4}}_{\vert\iia}\sp
R_i\vert_{\rm H}={\sqrt\frac{R_1 R_i}{ R_j R_k}}_{\vert\iia}\sp i,j,k=2,3,4 \ , 
\end{equation}
where the radii are measured in the respective string length units.
These relations are supplemented by
the identification of the string scale and six-dimensional string coupling,
\begin{equation}
\lh=\giis \lii\ \ ,\quad \giis=\frac{1}{\ghs} \ . 
\label{hetiia}
\end{equation}
The three-torus on which both sides are further reduced is inert
under heterotic-type II duality, so we easily find from \eqref{hetdil}
 the set of dilatonic
moduli in the Abelian part of the $SO(8,24)$ coset representative,
\begin{equation}
\label{iidil}
x_{i=1\dots 8}= \left( \sqrt{R_1 R_2 R_3 R_4},
\sqrt\frac{R_1 R_2}{ R_3 R_4}, \sqrt\frac{R_1 R_3}{ R_2 R_4},
\sqrt\frac{R_1 R_4}{ R_2 R_3},
\frac{R_5}{\giis},\frac{R_6}{\giis},\frac{R_7}{\giis},
\frac{R_5 R_6 R_7}{\giis} \right) \ , 
\end{equation}
where all radii are measured in units of the type II string length $\lii$.

The moduli appearing in \eqref{iidil} are not
the usual $SO(4,4)\times SO(3,3)$ that arise from the
reduction of the six-dimensional
type II theory on $T^4/\Zint_2 \times T^3$.
The latter can however be reached by defining
\begin{subequations}
\begin{eqnarray}
y_1 &=& \sqrt{x_1 x_2 x_3 x_4} \sp y_i = \sqrt\frac{ x_1 x_i}{ x_j x_k}
\sp i,j,k = 2,3,4 \\
y_8 &=& \sqrt{x_5 x_6 x_7 x_8} \sp y_i = \sqrt\frac{ x_8 x_i}{ x_j x_k}
\sp i,j,k = 5,6,7
\end{eqnarray}
\end{subequations}
which indeed gives the more familiar parameterization
\begin{equation}
\label{iidi2}
y_{i=1\dots 8}= \left( R_1 , \dots, R_7, 1/\giit^2 \right) \ .
\end{equation}
The type IIA theory compactified on $K_3\times T^3$ therefore also
appears to have a dynamically generated dimension, of size $\tilde R_8
=\lii/\giit^2$.
Again, this is not the same as the M-theory eleventh dimension of
size $R_{11}=\gii \lii$, nor is it identical to the 8th radius
$R_8=\lh/\ght^2=V_3/\lii^2$ that appeared naturally on the heterotic side.
We see that the mapping of heterotic/$T^4\times
T^3$ to type II on $K_3\times T^3$
involves both an $SO(4,4)$ triality on the $T^4/\Zint_2$ moduli, and
another $SO(4,4)$ triality on the non-perturbative $T^4$ torus.
This mapping amounts to a non-trivial embedding of the perturbative
$SO(4,20)\times SO(3,3)$ type II T-duality into the U-duality
group $SO(8,24)$. We have not worked out in detail the mapping
of the Borel part of the moduli, but this can be  easily done
along the lines of \cite{Kiritsis:2000zi}.

\section{Exact $F^4$ threshold in $D=3$}
Since $SO(8,24,\Zint)$ is believed to be an exact quantum symmetry
of the heterotic string compactified on $T^7$ or its dual versions,
all amplitudes should be invariant under this duality group.
In this section, we propose a U-duality invariant expression
for $F^4$ amplitudes, which reproduces the known
one-loop and tree-level answers on the heterotic and type II
sides respectively. By analyzing our proposal at weak coupling,
we will be able to identify the instanton configurations that
contribute, and extract their summation measure.

\subsection{U-dual completion of perturbative amplitudes}

Our starting point is the heterotic one-loop $F^4$ amplitude,
which already incorporates most of the
symmetries, namely the T-duality $SO(7,23,\Zint)$. It has been
discussed in great detail in \cite{Kiritsis:2000zi}, building on previous work
\cite{Lerche:1987sg}. The result is
\begin{equation}
\label{hetloop}
\Delta_{\rm 1-loop}= \lh^{5}
\int_{\cal F} \frac{d^{2}\tau}{\tau_2^2}  \frac{p_R^4}{\bar \eta^{24}} 
Z_{7,23} ~t_8 F^4  \ . 
\end{equation}
Here, $p_R$ has modular weight $(0,1)$, and inserts
the right-moving momentum corresponding to the choice of gauge boson $F$.
Dualizing the vectors into scalars, we find
\begin{equation}
\label{hetloops}
\Delta_{\rm 1-loop}= \lh \ght^8
\int_{\cal F} \frac{d^{2}\tau}{\tau_2^2}  \frac{p_R^4}{\bar \eta^{24}}
Z_{7,23} ~t_4 (M_{7,23}\p \phi)^4 \ , 
\end{equation}
where $t_4$ is the tensor $t_8$ with pairs of indices raised with the
$\e_3$ antisymmetric tensor.

It is now simple to covariantize this result under U-duality.
We propose that the exact $(\p \phi)^4$ threshold in heterotic
string on $T^7$, or any of its dual formulations, is given by
\begin{equation}
\label{uduala}
I_{(\p\phi)^4} = \lp \int d^3 x \sqrt{g} \int_{\cal F}
\frac{d^2\tau}{\tau_2^2} 
\frac{Z_{8,24}(g/\lh^2,b,\phi,\ght^2)}{\bar \eta^{24}}
~t_8 [  (e_{8,24}^{-1} \p_{\mu} e_{8,24})_{ia} ~ p_R^{a}] ^4 \ , 
\end{equation}
where $\lp = \ght^2 \lh$ is the three-dimensional (U-duality invariant)
Planck length. In this
expression,  $Z_{8,24}$ is the Theta function of the non-perturbative
$\Gamma_{8,24}$ lattice specified by \eqref{hetbein2},
\begin{subequations}
\begin{equation}
Z_{8,24}(\tau,\bar\tau)
= (\tau_2)^{4}  \sum_{m_i,p_I,n^i} e^{ -  \pi \tau_2 \M^2  - 2\pi
i \tau_1 v^t \eta_{8,24} v   }
\end{equation}
\begin{equation}
\M^2 = v^t M_{8,24} v \ ,\quad v =(m_i,p_I,n^i)\ ,\quad i
=1\ldots 8\ ,\quad I = 1 \ldots 16 \  . 
\end{equation}
\end{subequations}
where $M_{8,24}=e_{8,24}^t~ e_{8,24}$ is obtained from  \eqref{hetbein2},
and $\eta_{8,24}$ is the $SO(8,24)$ vielbein given in \eqref{eta}.
The momenta, windings and gauge charges $m_i, n^i, p^I$ are summed over the
even self-dual lattice $\Gamma_{1,1}^8\oplus D_{16}$.
$Z_{8,24}$ is invariant under both U-duality $SO(8,24,\Zint)$ and
$Sl(2,\Zint)$ modular transformations of $\tau$.
$e_{8,24}^{-1}\p_\mu e_{8,24}$ is the left-invariant one-form on
the coset $[SO(8)\times SO(24)]\backslash SO(8,24)$, pulled-back
to space-time. It takes
value in the $(\irrep{8},\irrep{24})$ component of the decomposition
of the Lie algebra $so(8,24)$ under $so(8)\times so(24)$.
Each index $a$ in the \irrep{24} is contracted with 
an insertion $p_R$ of the
right-moving momenta $(p_I,n^i)$ into the partition function $Z_{8,24}$.
The four indices $i$ in the \irrep{8} are contracted with the 
four momenta $\p_{\mu}$
using the tensor $t_8$. This structure is also the one
that arises in a one-loop four-scalar amplitude as shown
in \cite{Antoniadis:1998zt}.
The conjecture \eqref{uduala} satisfies the following criteria:
\begin{enumerate}
\item[(i)] It is $SO(8,24,\Zint)$ invariant by construction;
\item[(ii)] It correctly reproduces the heterotic 1-loop coupling;
\item[(iii)] The non-perturbative contributions come from heterotic
5-branes and KK5-branes, which are the expected ones in $D=3$;
\item[(iv)] The result decompactifies to the known purely perturbative
   result in $D \geq 4$;
\item[(v)] Via heterotic/type II and heterotic/type I
duality, the corresponding amplitude in type II and I shows
the correct perturbative terms and the expected instanton
corrections.
\end{enumerate}
We shall now proceed to prove these claims. In order to avoid
unnecessary complications, we will restrict ourselves to a
particular subspace of moduli space, corresponding to
the $T^4/\Zint_2\times T^3$ orbifold point on the type II side.
It has been shown in \cite{Lerche:1998nx,Kiritsis:2000zi} 
that for this choice of
Borel moduli, the Dedekind function $\bar \eta^{24}$
in \eqref{uduala} cancels against the action of $p_R^4$ on
the $D_{16}$ part of the lattice, and we are left with the
simpler expression
\begin{equation}
\label{np824}
\Delta_{(\p\phi)^4} =  \lp  \int_{\cal F}
\frac{d^2\tau}{\tau_2^2} Z_{8,8}(g/\lh^2,b,\ght^2) \ , 
\end{equation}
where the partition function $Z_{8,8}$ now runs over the lattice
$\Gamma_{1,1}^8$ only.

\subsection{The heterotic instanton expansion \label{hetiexp}}

To perform a weak coupling expansion of the result \eqref{np824} we
notice that defining $1/\ght^2 = R_8/\lh$ as in \eqref{r8def},
the weak coupling expansion becomes a large $R_8$ expansion,
so that we can adopt a
Lagrangian representation for the $S^1$ part and a Hamiltonian
representation for the remainder:
\begin{gather}
\label{del7lh} \Delta_{(\p\phi)^4}= \lp \frac{R_8}{\lh} \int_{\cal F}
\frac{d^{2}\tau}{\tau_2^2} \sum_{p,q} \sum_{v} \exp\left( -
\pi \frac {R_8^2|p-\tau q|^2 }{\lh^2 \tau_2} + 2\pi i p~ w_i n^i
\right) \tau_2^{7/2}
 q^{\frac{p_L^2}{2}} \bar q^{\frac{p_R^2}{2}} \ ,
\end{gather}
where $v=(m_i,n^i)$ now denotes the $7+7$ perturbative momenta and windings.
We apply
the standard orbit decomposition method on the two integers $(p,q)$,
trading the sum over $Sl(2,\Zint)$ images of $(p,q)$ for a sum
over images of the fundamental domain ${\cal F}$. The zero orbit
gives back the perturbative result \eqref{hetloop}
\begin{gather}
\label{del7z} \Delta_{(\p\phi)^4}^{\rm zero}=\lh
\int_{\cal F} \frac{d^{2}\tau}{\tau_2^2}  Z_{7,23} \ . 
\end{gather}
The degenerate orbit on the other hand, with representatives
$(p,0)$, can be unfolded onto the strip $|\tau_1|<1/2$. The
$\tau_1$ integral then imposes the level matching condition
$p_L^2-p_R^2=m_i n^i =0$, and the $\tau_2$ integral can be carried
out in terms of Bessel functions to give
\begin{equation}
\label{del7d} \Delta_{(\p\phi)^4}^{\rm deg}=2 \lh \sum_{p\neq 0}
\sum_{v \neq 0}  \delta(v^t \eta_{7,7} v ) \left( \frac{p^2}{\ght^4
v^t M_{7,7} v} \right)^{5/4} K_{5/2}\left( \frac{2 \pi}{\ght^2}
|p| \sqrt{v^t M_{7,7} v} \right) e^{2\pi i p w_i n^i}\ . 
\end{equation}
{}From the expansion of the Bessel function
\begin{equation}
\label{bes}
K_{5/2}(x) = \sqrt{\frac{\pi}{ 2x }} e^{-x} \left[ 1 + \frac{3}{x}
+ \frac{3}{x^2} \right]
\end{equation}
we see that these are non-perturbative contributions with classical action
\begin{equation}
\label{clah}
{\rm Re} (S_{\rm cl}) = \frac{2 \pi}{\ght^2}
|p| \sqrt{v^t M_{7,7} v} \ .  
\end{equation}
Choosing for $v$ either ``momentum'' charges or ``winding'' charges,
we find an action
\begin{equation}
\frac{1}{\ght^2} \frac{\lh}{R_i} = \frac{V_6}{\gh^2 \lh^6} \sp
\frac{1}{\ght^2}  \frac{R_i}{\lh} = \frac{V_6 R_i^2}{\gh^2 \lh^8} \ , 
\end{equation}
which identifies these instantons as heterotic 5 branes and KK5-branes
respectively, wrapped on a $T^6$ inside $T^7$. The summation measure
for these effects is easily extracted, and yields
\begin{equation}
\label{kk5m}
\mu_{\rm Het} (N) = \sum_{d|N} \frac{1}{d^5} \ , 
\end{equation}
where $N=gcd(p,m_i,n^i)$. We also note that the Bessel function $K_{5/2}$
in \eqref{bes}
exhibits only two subleading terms beyond the saddle point approximation,
so that these instanton contributions do not receive
any corrections beyond two loops.

Since NS5-brane instantons appear in the three-dimensional $(\p \phi)^4$
result, one may wonder how they can not contribute in four-dimensions,
where the $F^4$ threshold has been argued to be purely one-loop
\cite{Kiritsis:1999ss,Kiritsis:2000zi}.
Let us therefore investigate the decompactification of our result to
four dimensions, using the parameterization described at the end of
Subsection \ref{het7}. We thus
decompose $Z_{8,24} = Z_{2,2} Z_{6,6}$ where
$Z_{2,2}$ stands for the lattice sum on the
non-perturbative two torus in the (7,8) direction and $Z_{6,22}$
is the lattice sum for the perturbative states in $D=4$. From
\eqref{t2mod} we see that decompactifying to 4D corresponds
to a large volume  limit of the two-torus, so that we use a
Lagrangian representation for $Z_{2,2}$ and Hamiltonian for
$Z_{6,6}$,
\begin{gather}
\label{del7lh2} \Delta_{(\p\phi)^4}= \lp \frac{R_7 R_8}{\lh^2}
\int_{\cal F} \frac{d^{2}\tau}{\tau_2^2} \sum_{p^\alpha,q^\alpha}
\sum_{v}
 e^{ -\frac{\pi}{\tau_2} (p^\alpha + \tau
q^\alpha ) (g_{\alpha \beta}+B_{\alpha \beta}) (p^\beta + \bar
\tau q^\beta) + 2\pi i p^\alpha w_{\alpha j} n^j } \tau_2^3
q^{\frac{p_L^2}{2}} \bar q^{\frac{p_R^2}{2}} \ ,
\end{gather}
where $\alpha=7,8$ and $v$ are the perturbative charges in $D=4$.
We can now use the orbit
decomposition for the sum over $p,q$. Then, the trivial orbit
gives
\begin{gather}
\label{del7lh2z} \Delta^{{\rm zero};D=4}_{(\p\phi)^4}=  R_7
\int_{\cal F} \frac{d^{2}\tau}{\tau_2^2}
Z_{6,6} =
\frac{R_7}{\lh^4}  \Delta_{F^4}^{\rm 1-loop,4D}
\end{gather}
which decompactifies to the one-loop result in 4D. The degenerate orbit
has $q =0$, so that the $\tau_1$ integral imposes the
level-matching condition and the $\tau_2$ integral gives
\begin{gather}
\label{del7lh2d}
\Delta_{(\p\phi)^4}^{{\rm deg};D=4}= R_7 \sum_{p\neq 0} \sum_{v \neq 0}
\delta(v^t \eta_{6,6} v) \frac{|p^t g p |}{ |v^t M_{6,6} v | } K_2\left( 2
\pi \sqrt{|(p^t g p)( v^t M_{6,6} v)|} \right) e^{2\pi i p^t w n}\ . 
\end{gather}
Using the metric on the (7,8) torus
\begin{equation}
g =e_2^t e_2  \sp e_2 = \frac{R_7}{\lh} \begin{pmatrix} 1 & \tau_1
\\ 0 & \tau_2 \end{pmatrix}
\end{equation}
with $\tau$ defined in \eqref{t2mod}, one finds
$\sqrt{p^t g p} = \frac{R_7}{\lh} |p_1 + \tau  p_2 |$. It is then obvious
from the asymptotic form of the Bessel function
that the term \eqref{del7lh2d} disappears in 4D.
The non-degenerate orbit vanishes similarly in the limit $R_7
\rightarrow \infty$. We thus conclude that the conjecture \eqref{uduala}
correctly reduces to the perturbative result in $D \geq 4$.

\subsection{The type II  instanton expansion \label{iiiexp}}
We now consider the type II interpretation of our conjecture \eqref{uduala}.
{}From the moduli identification \eqref{iidil}, we see
that the weak coupling
expansion on the type II side corresponds to the large
volume expansion of the non-perturbative $T^4$ in the 5--8
directions, with volume
\begin{equation}
v_4 = \frac{V_3^2}{\giis^4 \lii^6}  \ , 
\end{equation}
where $V_3 = R_5 R_6 R_7$.
Using the by now familiar method, we thus decompose
$Z_{8,8} = Z_{4,4}^{(1-4)} Z_{4,4}^{(5-8)} $ with Hamiltonian and Lagrangian
representation for the two factors respectively, so that
\begin{gather}
\label{delii3} \Delta_{(\p\phi)^4}= \lp  v_4  \int_{\cal F}
\frac{d^{2}\tau}{\tau_2^2} \sum_{p^\alpha,q^\alpha} \sum_{m_i,n^i}
 e^{ -\frac{\pi}{\tau_2} (p^\alpha + \tau
q^\alpha ) (g_{\alpha \beta}+B_{\alpha \beta}) (p^\beta + \bar
\tau q^\beta) + 2\pi i p^\alpha w_{\alpha j} n^j } \tau_2^2
q^{\frac{p_L^2}{2}} \bar q^{\frac{p_R^2}{2}} \ ,
\end{gather}
where $\alpha=5,6,7,8$ and $i=1\ldots 4$. For use below, we define
the rescaled metric $G_4$ on the non-perturbative four-torus
\begin{equation}
\label{G4def}
G_4 = \giis^2 g_4 \equiv e_4^t e_4 \sp
e_4 = {\rm diag} (R_I/\lh,V_3/\lh^3) \sp I = 5,6,7 \ , 
\end{equation}
which depends on the geometric moduli only.

We can now perform in \eqref{delii3} the orbit
decomposition for the sum over $p,q$ (see
\cite{Kiritsis:1997em,Kiritsis:1997hf} for a discussion
of the orbit decomposition for $Z_{d,d}$, $d > 2$, which
generalizes the $d=2$ case \cite{Dixon:1991pc}.) Then, the trivial orbit
gives
\begin{gather}
\label{delii3t} \Delta_{(\p\phi)^4}^{\rm zero}= \giit^2 \lii
\frac{V_3^2}{\giis^4} \int_{\cal F} \frac{d^{2}\tau}{\tau_2^2}
Z_{4,4} =
 \frac{V_3}{(\giis \lii) ^4} \Delta_{F^4}^{\rm tree,6D} \ , 
\end{gather}
which, using \eqref{hetiia}
shows the correct 3D tree-level result,
directly induced from the 6D tree-level $F^4$ coupling
 \cite{Kiritsis:2000zi}.
The degenerate orbit,
with $q^\alpha =0$, can be unfolded onto the strip $|\tau_1|<1/2$. The
$\tau_1$ integral then imposes the level matching condition
$p_L^2-p_R^2=2m_i n^i=0$, and the $\tau_2$ integral can be carried
out in terms of Bessel functions to give
\begin{equation}
\label{delii3d} \Delta_{(\p\phi)^4}^{\rm deg}=2 \frac{V_3}{\giis^2
\lii^2} \sum_{p^i \neq 0} \sum_{(m_i,n^i)\neq 0}  \delta(m_i n^i)
\frac{1}{\giis} \frac{\sqrt{|p^t G_4 p |} }{ \sqrt{m^t M_{4,4} m} }
\end{equation}
$$ \;\;\;\;\;\;\;\;\;\;\;
\cdot K_1\left(  \frac{2\pi}{\giis} \sqrt{p^t G_4 p}  \sqrt{m^t
M_{4,4} m} \right) e^{2\pi i p^{\alpha} w_{\alpha i} n^i}\ . $$
{}From the argument of the Bessel function $K_1$,
we recognize the contributions of Euclidean D-branes wrapped
on an even cycle of $K_3$, times a one-cycle of $T^3$ (for
$p$ in the $5,6,7$ directions of the non-perturbative torus),
or the whole $T^3$ for $p$ in the 8th direction. The latter
case corresponds to the contributions of Euclidean D6-branes,
which start contributing in three dimensions.

For the non-degenerate orbit, the integral is dominated by the saddle point
\begin{subequations}
\begin{eqnarray}
q^{\alpha}  g_{\alpha \beta} (p^\beta-\tau_1 q^\beta)+ i \tau_2 m_i n^i&=&0 \ , \\
-(p^\alpha-\tau_1 q^\alpha)g_{\alpha \beta}(p^\beta-\tau_1 q^\beta)
+\tau_2^2(q^\alpha g_{\alpha \beta} q^\beta +
m^t M_{4,4}m)&=&0 \ ,
\end{eqnarray}
\end{subequations}
as in our previous work \cite{Kiritsis:2000zi}, Section 5.5.
This saddle point gives new non-perturbative contributions
  \begin{equation}
\Delta_{(\p\phi)^4}^{\rm n.d.}= \frac{4V_3}{\giis^2 \lii^2} 
 \sum_{p^i,q^i} \sum_{m_i,n^i}
\left(\frac{(q^2)^2+q^2 m^t M_{4,4}
m+(m_in^i)^2}{p^2q^2-(pq)^2}\right)^{3/4} K_{3/2} \left( \Re
S_{\rm cl} \right) e^{i \Im S_{\rm cl}} \ .
\end{equation}
to the $F^4$ threshold in type II, with classical action
\begin{equation}
\Re S_{\rm cl}=2\pi \sqrt{ \frac{p^2 q^2- (pq)^2}{(q^2)^2} \left(
\frac{(q^2)^2}{\giis^4 \lii^4}
 + \frac{q^2 m^t M_{4,4} m }{\giis^2 \lii^2} + (m_i n^i)^2 \right) } \ . 
\end{equation}
In this expression,
all inner products of $p,q$ vectors are taken with the metric $G_4$
defined in \eqref{G4def}.
In particular, setting $m_i=n^i=0$ and switching on one charge at a time
for simplicity, we identify these non-perturbative effects as coming
from NS5-brane instantons, with action
\begin{equation}
 \frac{R_I R_J}{ \giis^2 \lii^2}  = \frac{V_{K_3} R_I R_J}{\gii^2 \lii^6}
\end{equation}
and KK5-brane instantons, with action
\begin{equation}
 \frac{V_3 R_K}{ \giis^2 \lii^4} =\frac{V_{K_3} R_I R_J R_K^2 }{\gii^2 \lii^8}
\ . 
\end{equation}
The former already occur in four dimensions, but the latter are genuine
three-dimensional effects. For general values of the charges $p,q,m$,
we obtain contributions from boundstates of NS5 and KK5-branes
with D-branes. It is a simple matter to derive
the summation measure for $N$ KK5-brane instantons
in type IIA/$K_3 \times T^3$,
\begin{equation}
\mu_{\rm IIA} (N) = \sum_{d |N} \frac{1}{d^3} \ , 
\end{equation}
which is identical to the one derived for NS5-brane instantons in
on IIA/$K_3 \times T^2$ \cite{Kiritsis:2000zi}.

\section{Discussion}
In this work, we have proposed a novel exact amplitude for
three-dimensional string theories with 16 supersymmetries,
by completing the one-loop heterotic $F^4$ threshold
into a U-duality invariant result. Our result generalizes
the field theory analysis of \cite{Paban:1998qy}, who found
the exact $F^4$ amplitude in three-dimensional $SU(2)$
Yang-Mills theory with 16 supersymmetries, to a full string theory context.
In contrast to this work, we have not rigorously  proven
our proposal, for lack of knowledge of both
supersymmetry constraints in 3d supergravity and
harmonic analysis on Grassmannians. Many consistency
checks support our claim, however, including
the agreement with the type II tree-level result,
the $D=4$ decompactification limit and sensible instanton expansions.

\TABLE{

\begin{tabular}{|c||c|c|c|} \hline
     & IIA/$K_3\times T^n$ & Het/$T^n$ & I/$T^n$ \\ \hline \hline
D$p$ & 2 ($F^4$)		   & 	   & 
 \begin{tabular}{c} D1: 0 ($F^4$)\\D5: 1 ($R^2$), 5 ($F^4$) \end{tabular} 
                                                    \\ \hline
NS5  & 3 ($F^4$)	& 1 ($R^2$), 5 ($F^4$) & \\ \hline
KK5  & 3 ($F^4$)	& 5 ($F^4$)	& 5 ($F^4$) \\ \hline
\end{tabular}
\caption{Instanton contributions to $F^4$
and $R^2$ couplings in theories with 16 supersymmetries. The
entry denotes the exponent $r$ appearing in the
summation measure $\sum_{d|N} d^{-r}$.}}

Our result is particularly interesting on the heterotic
side, where such examples are scarce and non-perturbative effects little
understood. In addition to the heterotic five-brane instantons
already found in \cite{Harvey:1996ir} 
and the dual type I D5-brane
instantons \cite{Hammou:1999in},  
 we have exhibited the contributions of
KK5-instantons, which come into play for compactifications on 7-manifolds
with a $U(1)$ isometry. On the type II side, we have recovered the
familiar D-brane and NS5-instantons, together with the D6-brane
and KK5-monopoles peculiar to three-dimensional compactifications.
Although we have hardly mentioned it, the type I picture is very similar
to the heterotic one: Using heterotic/type I duality 
\cite{Polchinski:1996df}, the heterotic
one-loop amplitude reproduces the familiar disk and cylinder amplitudes,
together with D1-instanton effects. The heterotic 5-brane and KK5-brane contributions turn
into type I D5-branes and KK5-branes, and their summation measure
is unaffected by the duality. The table above summarizes the
instanton summation measures for all cases known so far.
It is a challenging problem to rederive the measures for the
NS5-brane and KK5-instantons. A not lesser challenge will be
to confront the long-eluded problem of two-dimensional
supergravities with their formidable affine symmetry.

\section*{Acknowledgments}

We thank Elias Kiritis for useful remarks and correspondence.
This work is supported in part by TMR network ERBFMRX-CT96-0045.


\providecommand{\href}[2]{#2}\begingroup\raggedright\endgroup
\end{document}